\documentclass[a4paper,11pt]{article}
\usepackage{pos}
\usepackage{caption}
\usepackage{subcaption}
\usepackage{blindtext}

\usepackage{lineno}

\title{Measurement of $\omega$ meson production in pp and \\p--Pb collisions at $\sqrt{s_{\scalebox{0.5}{NN}}}=5.02~$TeV with ALICE}
\ShortTitle{Production of $\omega$ mesons in pp and p--Pb collisions}

\author[a]{Nicolas Strangmann}

\onbehalf{for the ALICE collaboration}

\affiliation[a]{Institut für Kernphysik, Johann Wolfgang Goethe-Universität Frankfurt, Frankfurt, Germany}

\emailAdd{strangmann@stud.uni-frankfurt.de}

\abstract{

The ALICE experiment at the LHC investigates the properties of the hot and dense nuclear matter created in heavy-ion collisions. By comparing the particle production in pp and p--Pb collisions, possible nuclear initial state effects can be isolated.
Measurements of the $\omega$ meson $p_\text{T}$-spectra in pp and p--Pb collisions not only allow for a determination of the nuclear modification factor $R_\text{pPb}$, but also provide insight into the fragmentation process and serve as vital input for decay background simulations for direct photons.

In this contribution, measurements of the $\omega$ meson production in pp and p--Pb collisions at $\sqrt{s_{\scalebox{0.5}{NN}}}=5.02~$TeV are presented. This includes the signal extraction and various corrections of the $\omega$ meson yields, leading to their production cross sections and the first measured nuclear modification factor $R_\text{pPb}$ of the $\omega$ meson at LHC energies.
}

\FullConference{%
  11th International Conference on Hard and Electromagnetic Probes of High Energy Nuclear Collisions\\
  March 27-31, 2023\\
  Aschaffenburg, Germany
}

\begin{document}
\maketitle


\section{Motivation}

\noindent 
Measurements of particle production at the LHC can help to constrain parton distribution functions as well as fragmentation functions. 
While the production of pseudoscalar mesons ($\pi$, $\eta$, $K$) has been extensively studied in recent years \cite{ALICE_0.9TeV,ALICE_RpA,ALICE_8TeV,ALICE_Kaon,LHCb_RpA}, the production of vector mesons like the $\omega$ has been investigated less \cite{ALICE_7TeV_omega}, hampering theoretical models to better describe the production of vector mesons \cite{Theory_Omega_pls}. 
\noindent Moreover, measuring the production of $\omega$ mesons in pp and p--Pb collisions, one can probe Cold Nuclear Matter (CNM) effects. 
Modifications due to this cold medium can be quantified by the nuclear modification factor $R_\text{pPb}$, the ratio of the production cross section in p--Pb collisions to that in pp collisions scaled by the number of nucleons within the Pb-nucleus:
\vspace{-5mm}

\begin{align}
    R_\text{pPb} = \frac{1}{A_\text{pPb}} \frac{\text{d}^2\sigma_\text{pPb}/\text{d}p_\text{T}\text{d}y}{\text{d}^2\sigma_\text{pp}/\text{d}p_\text{T}\text{d}y}. \label{eq:RpA}
\end{align}
Finally, a precise measurement of the $\omega$ meson production serves as vital input for direct photon measurements, as the $\omega$ meson is the third largest contributor to the background of the photon measurement \cite{ALICE_direct_photon_contributions}, and therefore this measurement of $\omega$ mesons reduces the systematic uncertainties of direct photon analyses.

\section{Experimental setup and data sets}

\noindent ALICE as a multipurpose experiment is very well suited for the measurement of $\omega$ mesons. The $\omega$ mesons decay very close to the collision vertex ($c\tau=23~\text{f}m$) \cite{PDG} and can be reconstructed in ALICE via the dominant decay channel into three pions ($\omega\rightarrow\pi^+\pi^-\pi^0$). Charged pions can be tracked in the inner tracking system (ITS) and the time projection chamber (TPC). The neutral pions decay almost exclusively (99\,\%) into two photons \cite{PDG}. These are either measured with the electromagnetic calorimeter (EMCal) or via the reconstruction of $e^+e^-$ pairs in the TPC originating from their conversion in the detector material with the so-called photon conversion method (PCM).
The data presented here was collected with a minimum bias trigger requiring a coincident signal in both VZERO detectors on either side of the experiment.
In November 2016, ALICE recorded about 600 million p--Pb collisions, and in November 2017 about 900 million pp collisions at $\sqrt{s_{\scalebox{0.5}{NN}}}=5.02~$TeV.
The integrated luminosity inspected in this analysis in pp (p--Pb) collisions amounts to $\mathcal{L}=18.5~$n$b^{-1}$ ($0.3~$n$b^{-1}$).

\section{Analysis}

\subsection{Pion reconstruction}

\noindent The reconstruction of the $\omega$ meson relies on a pure sample of pions to increase the significance of the $\omega$ signal and therefore the $p_\text{T}$-reach of the measurement. 
Charged pions are selected from all reconstructed tracks based on their specific energy loss in the TPC (n$\sigma<3$). 
The decay photons measured in the EMCal and those reconstructed using PCM are combined to reconstruct neutral pions. This results in the three neutral pion reconstruction methods EMCal, PCM, and PCM-EMCal, in case one of the photons converts.
For the subsequent $\omega$ meson reconstruction only photon pairs with a reconstructed invariant mass close to the neutral pion mass are used.

\subsection{$\omega$ meson reconstruction}

\begin{figure}
     \centering
     \begin{subfigure}[b]{0.325\textwidth}
         \centering
         \includegraphics[width=\textwidth]{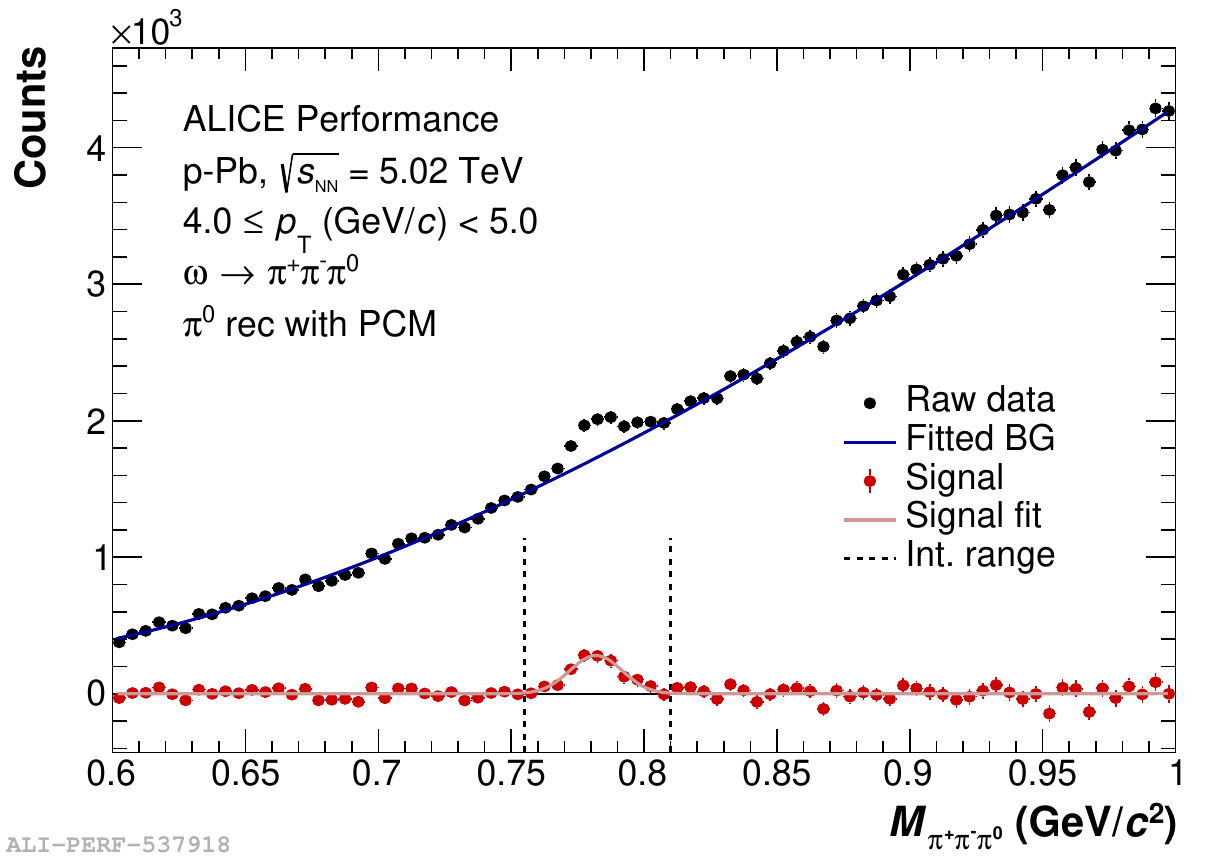}
         \caption{PCM}
         \label{fig:ExBinPCM}
     \end{subfigure}
     \hfill
     \begin{subfigure}[b]{0.325\textwidth}
         \centering
         \includegraphics[width=\textwidth]{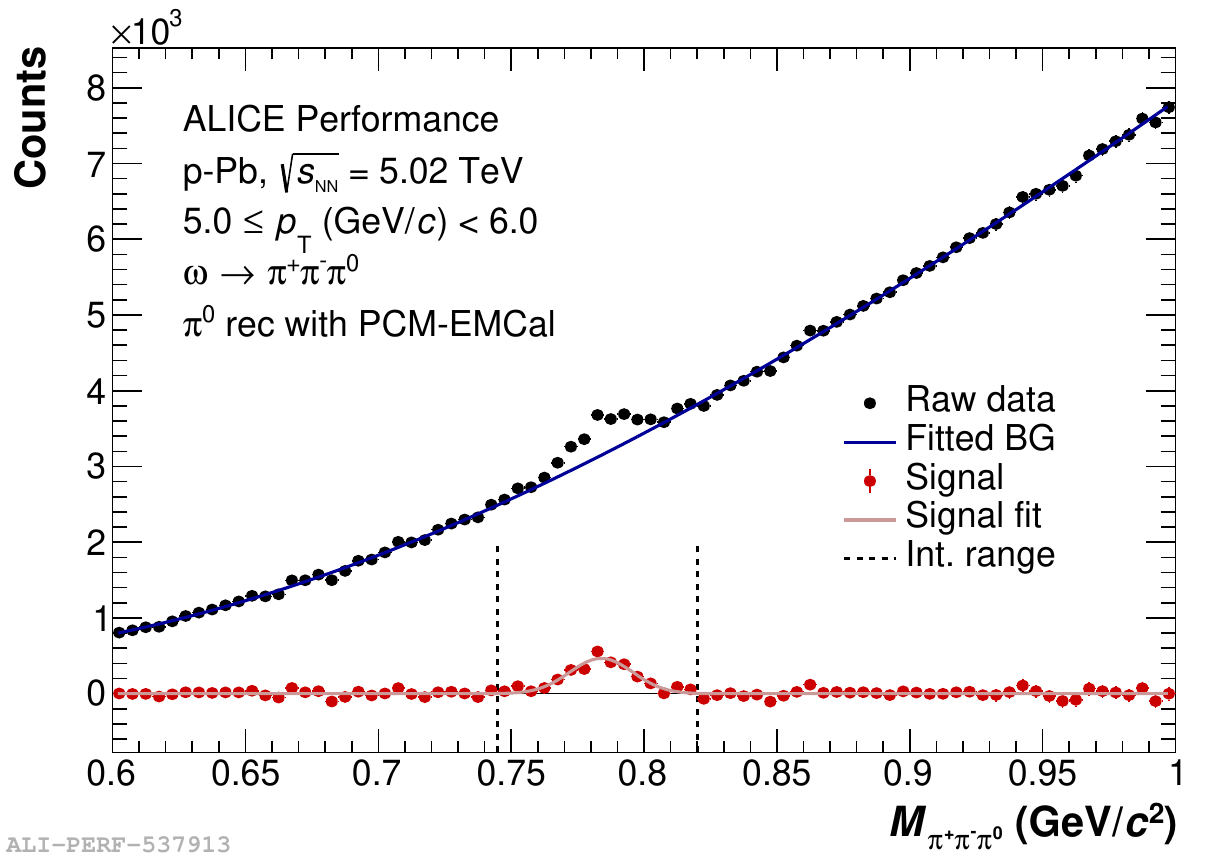}
         \caption{PCM-EMCal}
         \label{fig:ExBinPCMEMCal}
     \end{subfigure}
     \hfill
     \begin{subfigure}[b]{0.325\textwidth}
         \centering
         \includegraphics[width=\textwidth]{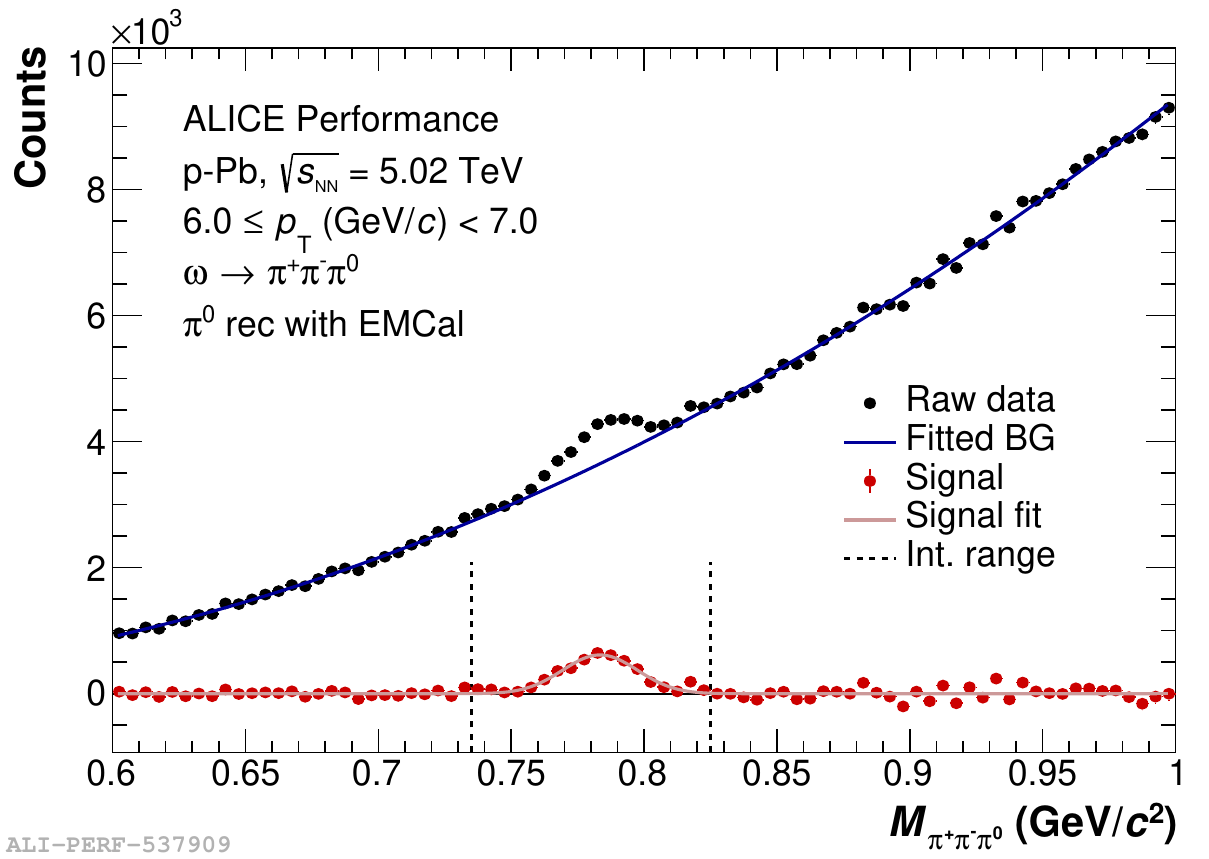}
         \caption{EMCal}
         \label{fig:ExBinEMCal}
     \end{subfigure}
        \caption{Distributions of the invariant mass of reconstructed $\omega$ meson candidates in p--Pb collisions using the three different $\pi^0$ reconstruction methods.}
        \label{fig:ExampleBins}
\end{figure}

\begin{figure}[h]
     \centering
     \begin{subfigure}[b]{0.49\textwidth}
         \centering
         \includegraphics[width=\textwidth]{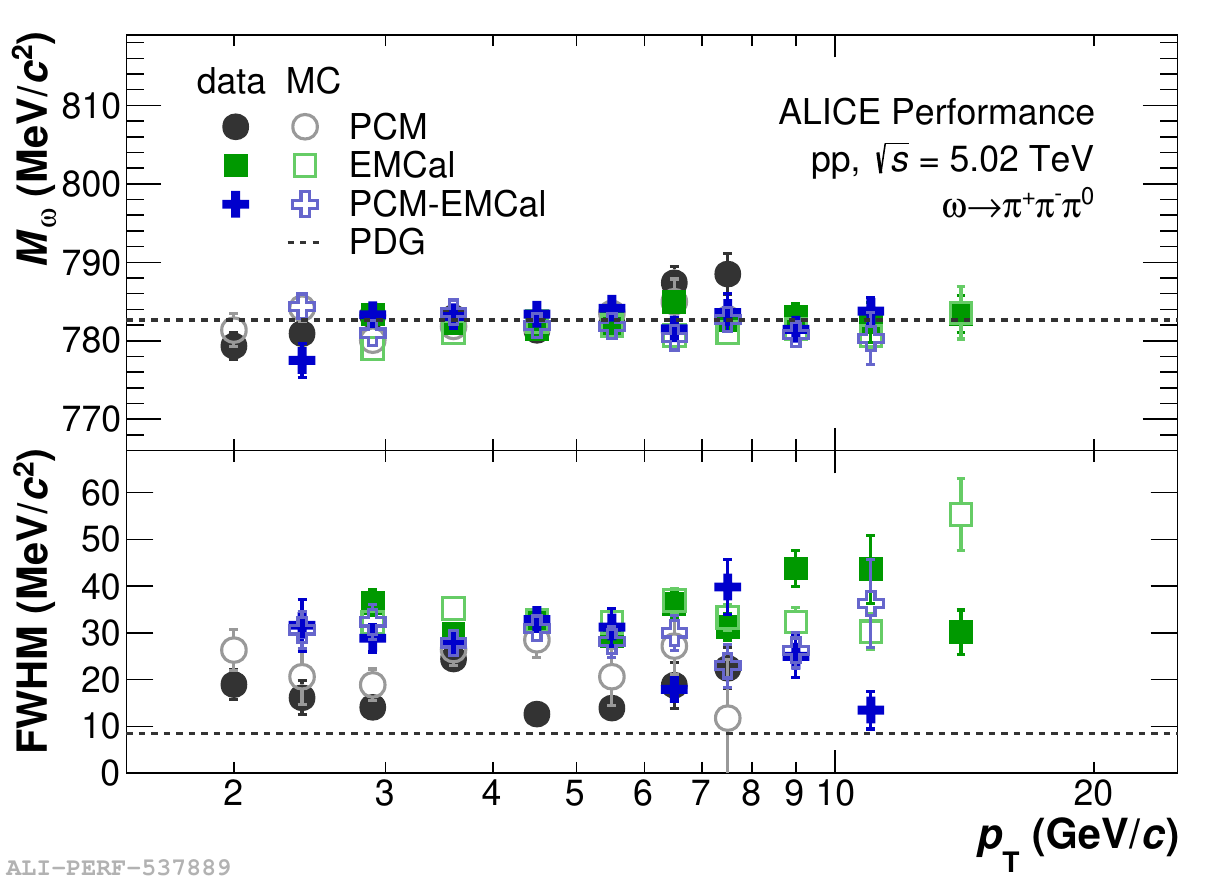}
         \caption{Mass and width in pp collisions}
         \label{fig:MethodRatiospp}
     \end{subfigure}
     \hfill
     \begin{subfigure}[b]{0.49\textwidth}
         \centering
         \includegraphics[width=\textwidth]{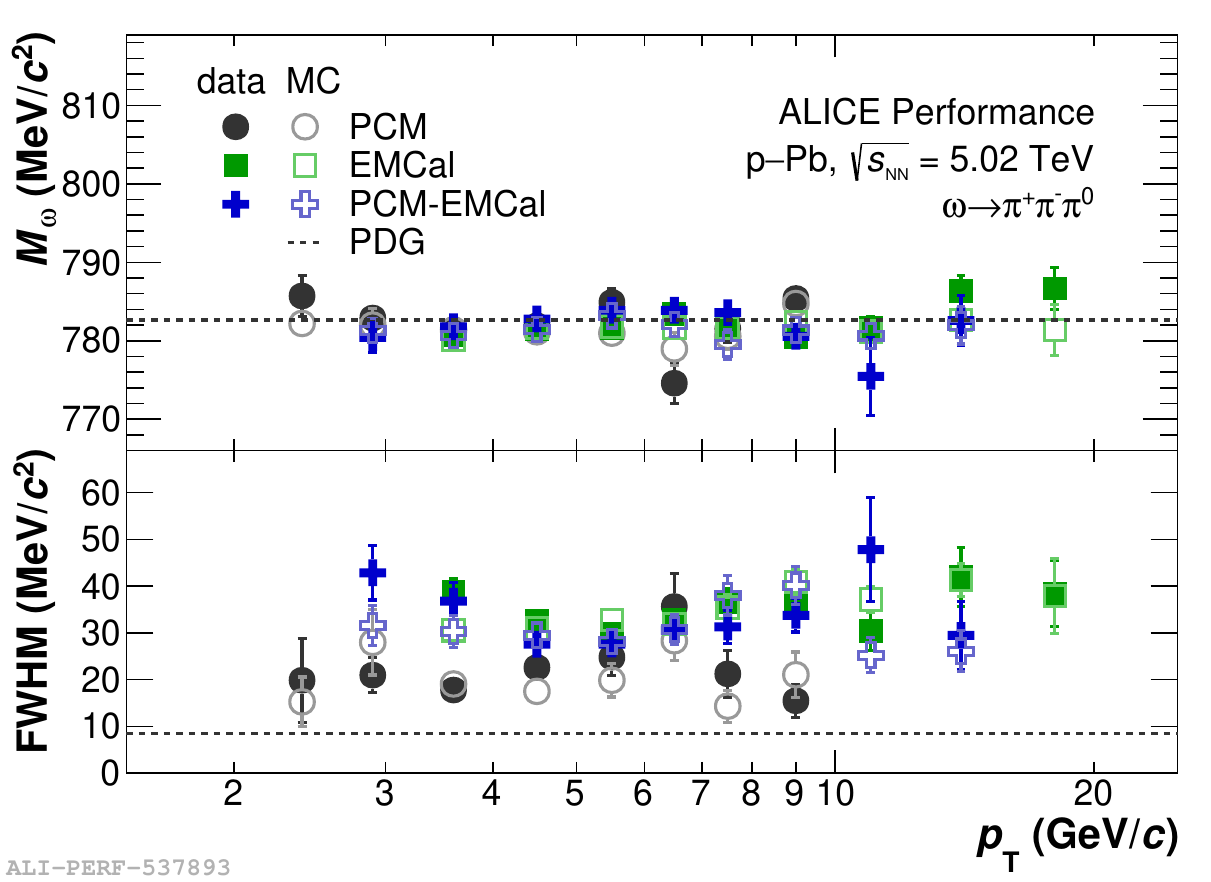}
         \caption{Mass and width in p--Pb collisions}
         \label{fig:MethodRatiospPb}
     \end{subfigure}
        \caption{Mass and width of the $\omega$ meson signal extracted using the different $\pi^0$ reconstruction methods. Full markers represent data, open markers the simulation, and the dotted lines show published values \cite{PDG}.}
        \label{fig:WidthMass}
\end{figure}

\noindent Combining all $\pi^+$, $\pi^-$ and $\pi^0$ mesons within each event, $\omega$ meson candidates are reconstructed. 
Figure \ref{fig:ExampleBins} shows the resulting number of $\omega$ meson candidates as a function of their reconstructed invariant mass in selected $p_\text{T}$ intervals in p--Pb collisions for the three neutral pion reconstruction methods.
\noindent
The combinatorial background can be described with a third order polynomial function. 
The $\omega$ meson signal, as shown in red in figure \ref{fig:ExampleBins}, is obtained by subtracting the estimated background.
The mass and width of the reconstructed $\omega$ mesons are extracted by parameterizing the signal with a Gaussian. The values obtained are shown in figure \ref{fig:WidthMass} for data (full markers) as well as for simulations (open markers). 
PYTHIA Monash 8.2 and DPMJET combined with GEANT3 were used to simulate the detector response.
While the reconstructed mass is in agreement with literature \cite{PDG}, the width is broadened by the detector resolutions. The extracted properties of the $\omega$ meson are well described by the Monte Carlo simulations.
The raw $\omega$ meson yield is extracted by integration of the signal around the reconstructed $\omega$ mass.

\subsection{Corrections and combination of methods}

\noindent
The Lorentz invariant production cross section is calculated from the extracted raw $\omega$ meson yield $N^{\omega}$ by applying corrections, as summarized in the following equation:

\begin{align}
    E \frac{\text{d}^3 \sigma_\omega}{\text{d}^3 p} = \frac{1}{\mathcal{L}_\text{int}} \frac{1}{2 \pi p_\text{T}} \frac{1}{A \epsilon \mathcal{B}} \frac{N^\omega}{\Delta p_\text{T}\Delta y}. \label{eq:XSec}
\end{align}

\begin{figure}
     \centering
     \begin{subfigure}[b]{0.49\textwidth}
         \centering
         \includegraphics[width=\textwidth]{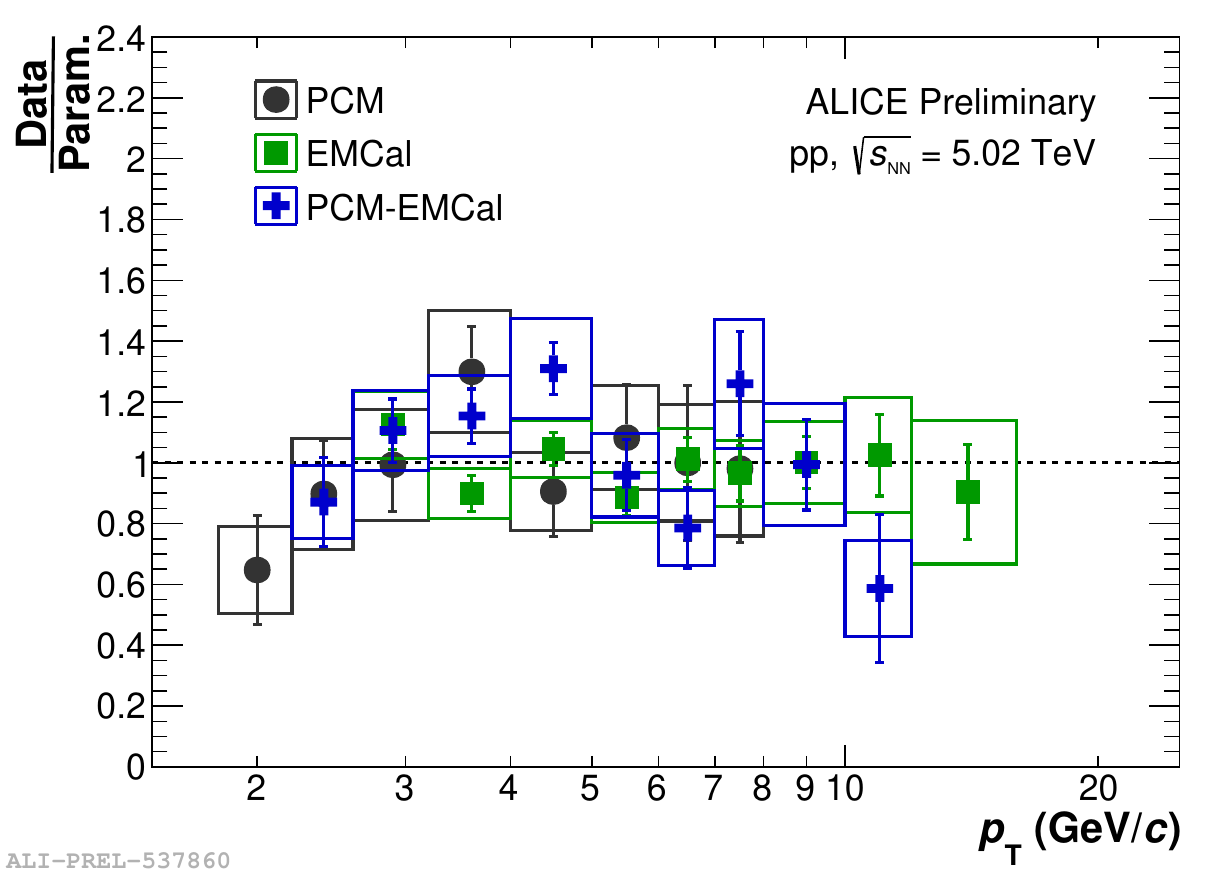}
         \caption{Ratios in pp collisions}
         \label{fig:MethodRatiospp}
     \end{subfigure}
     \hfill
     \begin{subfigure}[b]{0.49\textwidth}
         \centering
         \includegraphics[width=\textwidth]{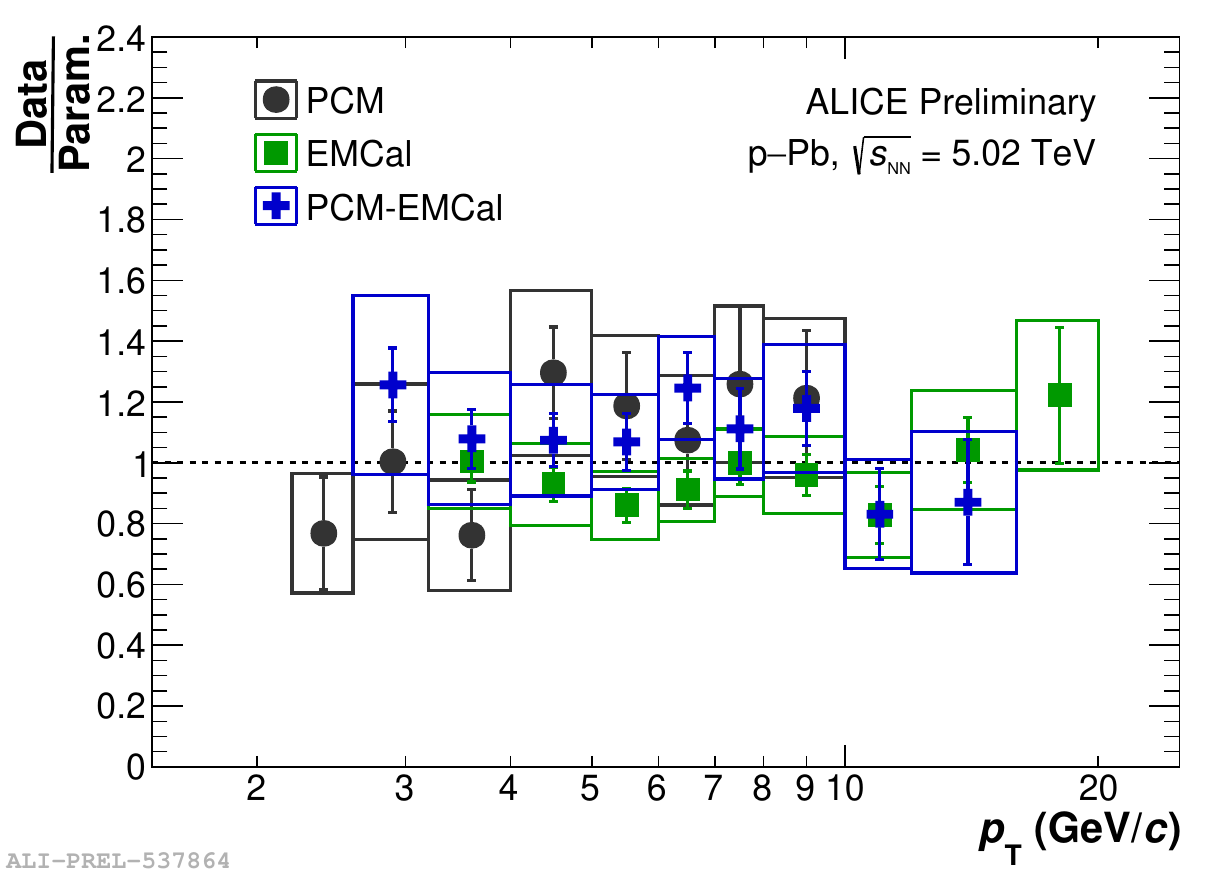}
         \caption{Ratios in p--Pb collisions}
         \label{fig:MethodRatiospPb}
     \end{subfigure}
        \caption{Ratios of the invariant cross sections extracted using the three different $\pi^0$ reconstruction methods to a combined Tsallis parametrization.}
        \label{fig:MethodRatios}
\end{figure}

\noindent 
Besides normalization by the Luminosity $\mathcal{L}_\text{int}$ and $2\pi \Delta y$, this includes the branching ratio $\mathcal{B}(\omega\rightarrow\pi^+\pi^-\gamma\gamma)\approx 89\%$ as well as the detector acceptance $A$ and reconstruction efficiency $\epsilon$. 
Using equations \ref{eq:XSec} and \ref{eq:RpA}, the cross sections in pp and p--Pb as well as the $R_\text{pPb}$ were calculated individually for each $\pi^0$ reconstruction method.
To account for the difference in the investigated rapidity regions in the two collision systems (see figure \ref{fig:XSec}), the pp spectrum used for the $R_\text{pPb}$ calculation was corrected by a factor of about $0.99$ based on the ratio of $\omega$ spectra generated with PYTHIA Monash 8.2 for both kinematic regions.
For the three reconstruction methods of the $\pi^0$ and thus $\omega$, the systematic uncertainties are estimated from variations of the applied pion selection criteria and extraction parameters. 
Figure \ref{fig:MethodRatios} shows the ratio of the extracted invariant cross sections to a Tsallis parametrization in pp and p--Pb. A good agreement is observed between these methods within the statistical (bars) and systematic (boxes) uncertainties.
The cross sections and nuclear modification factors extracted using the different methods are combined based on their uncertainties and correlations, using the Best Linear Unbiased Estimate (BLUE) method \cite{BLUE}.
While PCM allows for an extraction of the $\omega$ down to low $p_\text{T}$ thanks to its good energy resolution, the photon conversion probability of around 8\% restricts the range at higher transverse momenta, where the EMCal measurement dominates the measurement.

\section{Results}
\paragraph{Production cross section}

\begin{figure}
    \centering
    \begin{subfigure}[b]{0.555\textwidth}
          \includegraphics[width=\textwidth]{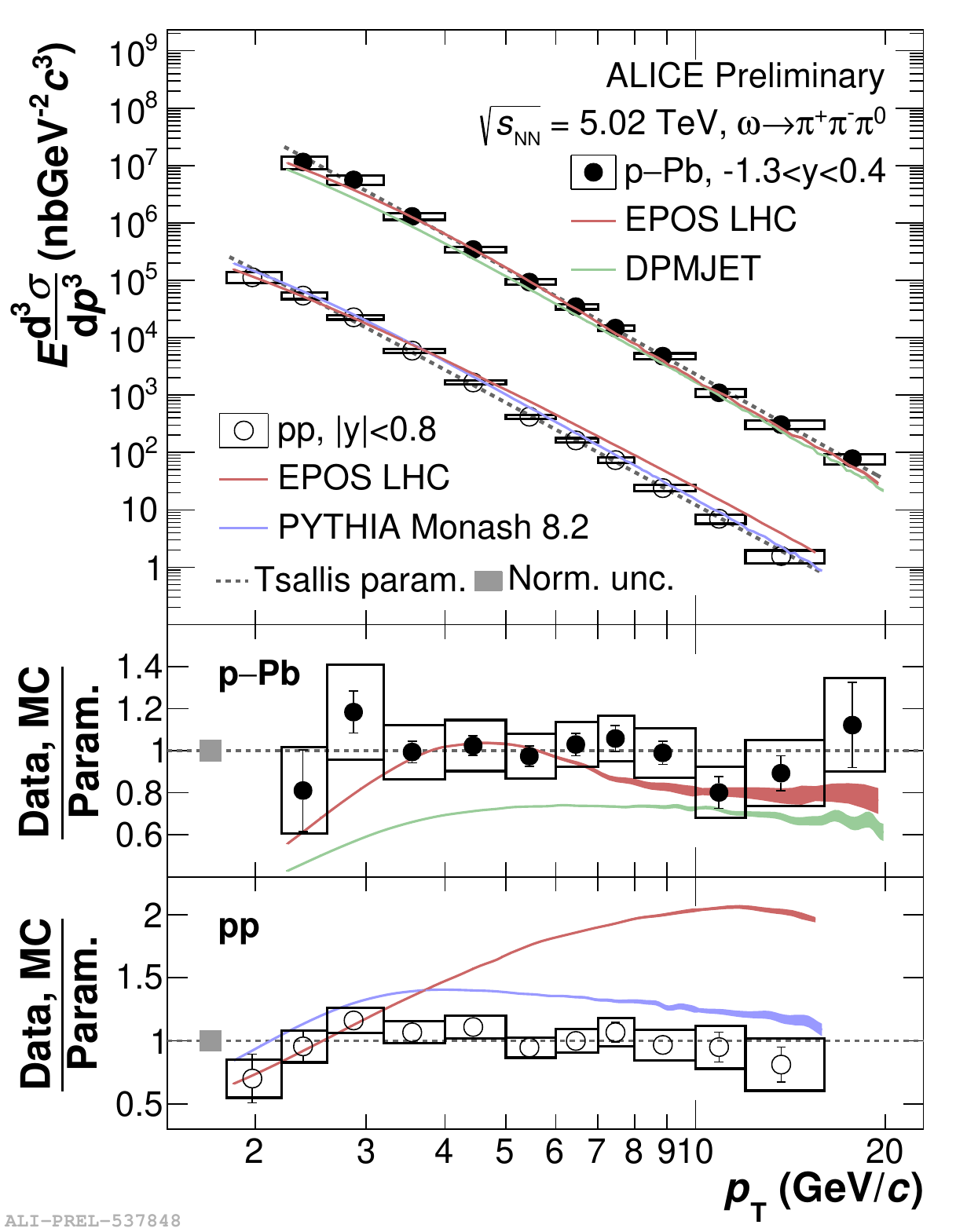}
      \caption{Measured production cross section of the $\omega$ meson in pp\\ and p--Pb collisions and selected MC generator predictions}\label{fig:XSec}
    \end{subfigure}
    \hfill
    \begin{minipage}[b]{0.435\textwidth}
      \begin{subfigure}[b]{\linewidth}
            \includegraphics[width=\textwidth]{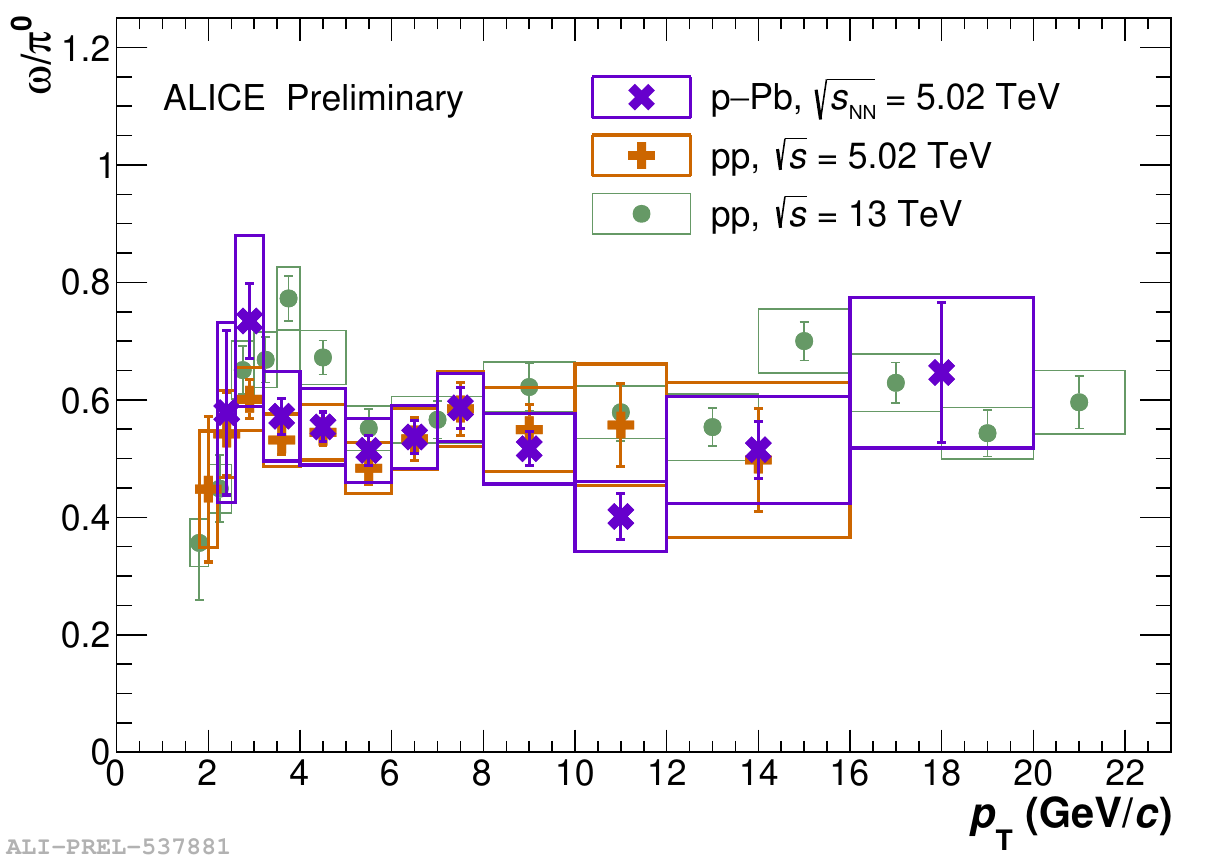}
        \caption{$\omega/\pi^0$ ratio measurements}\label{fig:OTP}
      \end{subfigure}\\[\baselineskip]
      \begin{subfigure}[b]{\linewidth}
            \includegraphics[width=\textwidth]{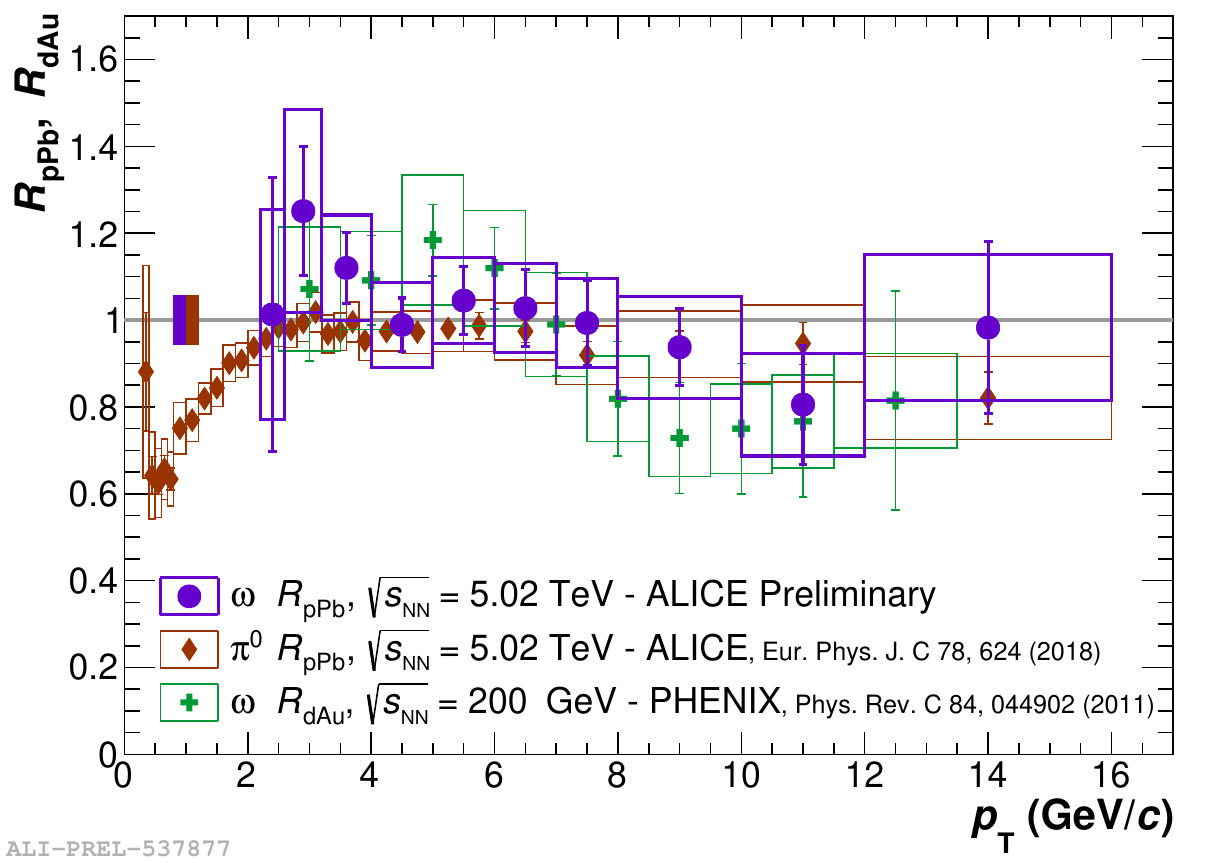}
        \caption{$\omega$ $R_\text{pPb}$ compared to $\pi^0$ $R_\text{pPb}$ \cite{ALICE_RpA} and $\omega$ $R_\text{dAu}$ \\at $\sqrt{s_{\scalebox{0.5}{NN}}}$\,=\,200\,GeV \cite{PHENIX_RpA}}\label{fig:RpA}
      \end{subfigure}
    \end{minipage}
    \caption{Measurements of $\omega$ meson production at $\sqrt{s_{\scalebox{0.5}{NN}}}=5.02~$TeV.}\label{fig:Results}
  \end{figure}

The extracted production cross sections of $\omega$ mesons for pp and p--Pb collisions at $\sqrt{s_{\scalebox{0.5}{NN}}}=5.02~$TeV are shown in figure \ref{fig:XSec}.
The pp cross section covers the transverse momentum range $1.8\le p_\text{T}<16\,\text{GeV}/c$ within $\vert y\vert < 0.85$. In p--Pb collisions, the cross section is extracted for transverse momenta $2.2\le p_\text{T}<20\,\text{GeV}/c$ for $-1.315<y<0.385$.
The lower panel in figure \ref{fig:XSec} shows the ratios of the data as well as various MC predictions to the respective Tsallis parametrization. While EPOS LHC describes the spectrum in p--Pb well, it overpredicts the production in pp by up to 100\,\%. 
PYTHIA Monash 8.2 and DPMJET describe the shape of the spectra in pp and p--Pb collisions, respectively, but over-/underestimate them by 30-40\,\%.

\paragraph{\boldmath{$\omega/\pi^0$} ratio}
The ratio of $\omega$ mesons to neutral pions as function of $p_\text{T}$ at $\sqrt{s_{\scalebox{0.5}{NN}}}=5.02~$TeV is shown in figure \ref{fig:OTP} for pp (orange) and p--Pb (violet) collisions. The result of a measurement of $\omega$ production at $\sqrt{s}=13~$TeV is shown in green. The agreement between the three measurements suggest an independence of the $\omega/\pi^0$ ratio of the collision system and energy within the uncertainties.

\paragraph{Nuclear modification factor}
Figure \ref{fig:RpA} shows the first measurement of the nuclear modification factor $R_\text{pPb}$ of the $\omega$ meson at LHC energies. As $R_\text{pPb}$ is consistent with unity, no nuclear modification is visible over the measured $p_\text{T}$ range. The measurement of $\pi^0$ $R_\text{pPb}$ \cite{ALICE_RpA} and $\omega$ $R_\text{dAu}$ at $\sqrt{s_{\scalebox{0.5}{NN}}}$\,=\,200\,GeV are both in agreement with the nuclear modification factor presented here within uncertainties.

\newpage
\section{Conclusion}
\noindent The $\omega$ meson production cross section was measured in pp and p--Pb collisions at $\sqrt{s_{\scalebox{0.5}{NN}}}=5.02~$TeV. \\
An agreement between the corresponding $\omega/\pi^0$ ratios in pp and p--Pb collisions and a previous measurement at $\sqrt{s}=13~$TeV suggest the production ratio to be independent of the collision system and energy.
A first measurement of the nuclear modification factor of the $\omega$ mesons presented in this contribution shows no visible nuclear modification over the measured $p_\text{T}$ range.\\

This contribution was supported by BMBF and the Helmholtz Association.

\end{document}